\documentstyle[preprint,pra,aps]{revtex}
\begin{document}
\draft
\title{
Comment on ``Consistency, amplitudes and 
probabilities in quantum theory'' by A. Caticha}
\author{Yoel Tikochinsky
\thanks{On leave from Racah Institute of Physics, The Hebrew University of Jerusalem.}
\thanks{Present address: Cavendish Laboratory, Madingley Road, Cambridge, CB3 0HE, U.K.}
}
\address{Department of Physics and Astronomy,
University of Canterbury, Private Bag 4800,
Christchurch, New Zealand}
\date{\today}
\maketitle
\begin{abstract}
A carefully written paper by A. Caticha [Phys. Rev. \textbf{A57}, 1572
(1998)] applies consistency arguments to derive the quantum mechanical rules
for compounding probability amplitudes in much the same way as earlier work
by the present author [J. Math. Phys. \textbf{29}, 398 (1988) and Int. J.
Theor. Phys. \textbf{27}, 543 (1998)]. These works are examined together to
find the minimal assumptions needed to obtain the most general results.
\end{abstract}


In a recent article \cite{AC98}, A. Caticha uses consistency arguments to derive the
quantum mechanical rules for combining probability amplitudes. Caticha's
work bears close resemblance, both in approach and execution, to earlier
work by the present author \cite{YT88a,YT88b}. With hindsight, it seems a good time to
take stock and see what has been achieved by this approach and what are the
minimal assumptions needed to obtain the most general results. Our point of
departure is the recognition that in quantum mechanics one cannot directly
assign probabilities to processes. In contrast to the classical situation,
not every proposition can be answered by yes or no (which slit did the
particle go through?). Therefore Boolean algebra does not apply and the road
opened by R.T. Cox \cite{RTC46} to introduce probabilities is not open to us.
Probability must, therefore, be introduced indirectly as a function of the
corresponding probability amplitude \cite{YT88b}. Let us review briefly how this is
done.

The basic entities of concern are the transition amplitudes $\langle B\mid
A\rangle $ between experimentally determined initial and final states A and
B. Both time dependent transitions $\langle B\left( t_{2}\right) \mid \left(
A(t_{1}\right) \rangle $ and transitions at a given time $t_{2}=\,t_{1}$ are
of interest. To each transition one assigns a complex number - the
probability amplitude for the process. This number is assumed to depend only
on the given process $A\rightarrow B$ and to be independent of the past
history (Markovian property).

Among the possible processes there are two kinds of special interest: 
\textit{processes in series}, an example of which will be $C\rightarrow
B\rightarrow A$ with amplitude $\langle A\mid C\rangle _{\text{via }B}$,
where $C$ is made to pass through a filter $B$ before $A$ is verified, and 
\textit{processes in parallel}, the simplest of which will be
$$
B{\nearrow \atop \searrow}
{C_{1} \atop C_{2}}
{\searrow \atop \nearrow }\;A
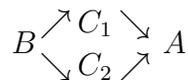
$$, where $B$ can proceed to $A$ only through
two orthogonal filters $C_{1},C_{2}$. The amplitude for this process will be
denoted by $\langle A\mid B\rangle _{\text{via }C_{1},C_{2}}.$ Very special
cases of these ``in series'' or ``in parallel'' processes are referred to as
AND, OR setups in [1]. Our \textit{only} assumption regarding these
processes is that the amplitudes for the processes are given analytic
functions of the partial complex amplitudes $x$ and $y$, namely, 
\begin{equation}
\langle A\mid C\rangle _{\text{via }B}=f(x,y)
\end{equation}
where 
\begin{equation}
x=\langle A\mid B\rangle \text{\quad and\quad }y=\langle B\mid C\rangle 
\text{,}
\end{equation}
and 
\begin{equation}
\langle A\mid B\rangle _{\text{via }C_{1}C_{2}}=g(x,y)\text{\quad }
\end{equation}
where 
\begin{equation}
x=\langle A\mid B\rangle _{\text{via }C_{1}}\text{\quad and\quad }y=\langle
A\mid B\rangle _{\text{via }C_{2}}.
\end{equation}
Our task is to find the possible form of these functions, subject to
consistency demands.
Consider now the process $D\rightarrow C\rightarrow B\rightarrow A$ with
amplitudes 
\begin{equation}
z=\langle C\mid D\rangle \text{,\quad }y=\langle B\mid C\rangle \quad \text{%
and }x=\langle A\mid B\rangle .
\end{equation}
We could calculate the amplitude for the process in two different ways: (a)
combine first $C\rightarrow B\rightarrow A$ to obtain $\langle A\mid
C\rangle _{\text{via }B}=$ $f(x,y)$ and then calculate $f(f(x,y),z)$ or, (b)
combine first $D\rightarrow C\rightarrow B$ to obtain $\langle B\mid
D\rangle _{\text{via }C}=$ $f(y,z)$ and then calculate $f(x,f(y,z))$.
Consistency then demands that the two calculations give the same result,
namely, the function $f(x,y)$ should obey the associative law 
\begin{equation}
f(x,f(y,z))=f(f(x,y),z).
\end{equation}
Similarly, for processes in parallel, consistency entails 
\begin{equation}
g(x,g(y,z))=g(g(x,y),z).
\end{equation}
Finally, consider the combined process 
$$
C\rightarrow B{\nearrow \atop \searrow }
{C_{1}  \atop C_{2}}
{\searrow\atop\nearrow }\;A\ .
$$
One way to calculate the amplitude is to
consider it as a process in series, with amplitude $f(g(x,y),z),$ where 
\begin{equation}
x=\langle A\mid B\rangle _{\text{via}\,C_{1}}\text{,\quad }y=\langle A\mid
B\rangle _{\text{via }C_{2}}\text{\quad and\quad }z=\langle B\mid C\rangle 
\text{ .}
\end{equation}
Another way to look at the same process is to consider it as a process in
parallel, with an amplitude $g(f(x,z),\,f(y,z)).$ Demanding that the two
representations agree, we have the distributive law 
\begin{equation}
f(g(x,y),z)=g(f(x,z),\text{ }f(y,z)).
\end{equation}

This is \textit{all }that is needed. From here on the rest is mathematics.
In particular, there is no need to assume commutativity for processes in
parallel, as was done in [1] and [2]. The equality $g(x,y)=g(y,x)$ follows
automatically from Cox's solution [5] of the associative law (7), as
recounted for example, in [1]. Disposing with commutativity renders the use
of artificial time-dependent filters [1] unnecessary, and allows general
formulation in terms of arbitrary filters and states. As shown in [2] and
[1], given the functions $g(x,y)$ and $f(x,y)$ it is always possible to find
a transformation $x^{\prime }=H(x)$ which will bring $g$ and $f$ to the 
\textit{canonical }form 
\begin{equation}
\lbrack g(x,y)]^{\prime }=x^{\prime }+y^{\prime }\text{,\quad }[f(x,y)%
]^{\prime }=x^{\prime }y^{\prime }\text{ .}
\label{eq10}
\end{equation}
Conversely, starting with Eq.~(\ref{eq10}), one can make a transformation $x^{\prime
\prime }=K(x^{\prime })$ such that, in terms of the new variables $x^{\prime
\prime }$ and $y^{\prime \prime }$, the addition and multiplication laws
(\ref{eq10}) change their \textit{form}, without changing their \textit{contents}.

From here on we shall restrict our discussion to \textit{transitions at a
given time}. Our aim is to amend the general proof of Born's law 
\begin{equation}
\Pr (A\mid B)=\mid x\mid ^{2},\quad x=\langle A\mid B\rangle
\end{equation}
given in [3]. Here $\Pr (A\mid B)$ stands for the probability of transition,
at a given time, from $B$ to $A$. To achieve this, neither dynamics nor the
introduction of dubious multi-particle filters [1] are needed. As shown in
[3], the amplitude for the inverse transition $A\rightarrow B$ satisfies $%
\langle B\mid A\rangle =\langle A\mid B\rangle ^{*}$. Futhermore, the
probability for the process $B\rightarrow A$ was shown there to be of the
form 
\begin{equation}
\Pr (A\mid B)=\mid x\mid ^{\alpha },\quad \alpha >0.
\end{equation}
Consider now all the orthogonal states $A_{i}$, which can be reached from $B$%
, with an amplitude $x_{i}=\langle A_{i}\mid B\rangle $. Since 
\begin{equation}
\langle B\mid B\rangle =\sum_{i}\langle B\mid A_{i}\rangle \langle A_{i}\mid
B\rangle =\sum_{i}x_{i}^{*}x_{i},
\end{equation}
and since the probability of the certain event satisfies $\Pr (B\mid B)=1$,
we have by (12) and (13) 
\begin{equation}
\Pr (B\mid B)=(\sum_{i}\mid x_{i}\mid ^{2})^{\alpha }=1.
\end{equation}
Hence, taking the logarithm of both sides, we obtain 
\begin{equation}
\sum \mid x_{i}\mid ^{2}=1.
\end{equation}
But the totality of processes $B\rightarrow A_{i}$ form an exhaustive and
mutually exclusive set of alternatives, satisfying (see Eq. (12)) 
\begin{equation}
\sum \Pr (A_{i}\mid B)=\sum \mid x_{i}\mid ^{\alpha }=1.
\end{equation}
Comparing (15) and (16) we find $\alpha =2$ and 
\begin{equation}
\Pr (A\mid B)=\mid x\mid ^{2}\text{ .}
\end{equation}

In summary, the assumptions (a) that amplitudes for processes in series or
in parallel are represented by analytic functions of the complex partial
amplitudes, and (b) that the probability of a process is a function of the
amplitude for the process, are enough to derive the known quantum mechanical
rules for combining amplitudes and for calculating the corresponding
probabilities. This is achieved using general states and filters. That these
assumptions are all that is needed, was not \textit{fully} realised either
in [2], [3] or in [1]. The work of Caticha certainly helped to put things in
sharper focus. In particular, as shown by Caticha, assumption (a) is enough
to establish the linearity of the Schr\"{o}dinger equation.

\begin{center}
\textbf{ACKNOWLEDGEMENT}
\end{center}
I would like to thank Geoff Stedman for drawing my attention to Ref. \cite{AC98}, and
for reading and commenting on this work.

\end{document}